\documentclass[a4paper,12pt]{article}

\usepackage[linkcolor=blue]{hyperref}
\usepackage{graphicx}
\usepackage{color}
\usepackage{url}
\usepackage{subfig}
\usepackage{authblk}

\begin{document}

\newcommand{\TODO}{\textcolor{red}{\bf TODO}\\}
\newcommand{\CITE}[1]{\textcolor{magenta}{\bf CITE}(\textcolor{blue}{#1})}
\newcommand{\GRC}{G_R}
\newcommand{\lrc}{c_R}
\newcommand{\pjump}{p_\mathrm{jump}}
\newcommand{\G}{\mathcal{G}}
\newcommand{\diff}{\mathrm{d}}
\newcommand{\cat}[1]{\emph{#1} }

\title{Universal hierarchical behavior of citation networks}
\author[1]{Enys Mones}
\author[2,3]{P\'{e}ter Pollner}
\author[1,2,3]{Tam\'{a}s Vicsek}
\affil[1]{Department of Biological Physics, E\"{o}tv\"{o}s Lor\'{a}nd University, P\'{a}zm\'{a}ny P\'{e}ter stny. 1/A, H-1117 Budapest, Hungary}
\affil[2]{MTA-ELTE Statistical and Biological Physics Research Group, P\'{a}zm\'{a}ny P\'{e}ter stny. 1/A, H-1117 Budapest, Hungary}
\affil[3]{ELTE Faculty of Sciences, Regional Knowledge Centre Ir\'{a}nyi D\'{a}niel u. 4, H-8000 Sz\'{e}kesfeh\'{e}rv\'{a}r, Hungary}
\date{\today}
\maketitle

\begin{abstract}
Many of the essential features of the evolution of scientific research are imprinted in the structure of citation networks.
Connections in these networks imply information about the transfer of knowledge among papers, or in other words, edges describe the impact of papers on other publications.
This inherent meaning of the edges infers that citation networks can exhibit hierarchical features, that is typical of networks based on decision-making.
In this paper, we investigate the hierarchical structure of citation networks consisting of papers in the same field.
We find that the majority of the networks follow a universal trend towards a highly hierarchical state, and i) the various fields display differences only concerning their phase in life (distance from the ``birth'' of a field) or ii) the characteristic time according to which they are approaching the stationary state.
We also show by a simple argument that the alterations in the behavior are related to and can be understood by the degree of specialization corresponding to the fields.
Our results suggest that during the accumulation of knowledge in a given field, some papers are gradually becoming relatively more influential than most of the other papers.
\end{abstract}

\section{Introduction}
\label{sec:introduction}

Citation networks are one of the simplest representations of the development of human knowledge, as they describe the connections and influences among academic papers in different fields \cite{garfield55,price65}.
Due to their importance, these networks have been the subject of various studies, including the distribution of citations \cite{price76}, community structures \cite{lancichietti10,chen10}, topology \cite{karrer09} or the history of single papers \cite{wang13}.
The available databases have shown spectacular growth recently, resulting in the access to detailed bibliographic data dating back even for decades.
These data provide the opportunity for the accurate study of the structural evolution of academic publications to the accessible period of time \cite{leicht07,eom11}.

Connections in a citation network are inherently asymmetric: in most cases the citation is an act solely of the citing paper.
There are, however, several reasons behind the inclusion of a publication in the reference list \cite{macroberts96} and there are also signs for cases in which the cited paper is not even read by the authors before citing it.
Nevertheless it is reasonable to assume, that a reference from paper $B$ to paper $A$ is related to some level of influence of paper $A$ on paper $B$.
In other words, the content of paper $B$ is affected by that of paper $A$ (at least in the choice of subject).
Through the life of particular papers, the network they form is permanently changing in time, resulting in a continuously developing structure.
There are different approaches in the literature to characterize this evolution: by the distribution of times a paper is cited \cite{redner98}; the number of cited references \cite{vazquez01}; or the changing in the number of papers in time \cite{price63} etc.
Most of these studies usually focus on local properties of the network, i.e., the small neighborhood of the nodes is considered, and they do not take into account the topology of the graph.
However, the global structure of the network itself opens the door to the investigation of other properties, which span over the whole system and describe a global behavior.

Hierarchy is a global and unique structural organization of many networks, and it is present in a wide range of systems: animal groups \cite{nagy10}, among employees of a company \cite{rowe07} and in social and technical networks \cite{pumain06}.
In directed networks, \emph{flow hierarchy} is defined as a layering of the nodes in which the direction of the edges obeys a ``global flow'', e.g., edges point from upper levels towards lower levels \cite{lane06}.
In systems of human-human interactions, the meaning of flow hierarchy is most pronounced in networks of decision-makings, i.e., where an edge pointing from person $A$ to person $B$ represents the fact that the decision made by member $A$ is taken into consideration by $B$ while making its own decision.
It has been shown that in a general framework, where greedy agents are supposed to solve consecutive tasks and optimize their success rate, a decision-following hierarchy emerges, if they are provided by the answers of other agents \cite{nepusz13}.
Citation networks can be considered as an abstract decision-following network among scientific publications, and an occurring hierarchical feature would suggest that the phenomenon of hierarchy is 
not restricted to small groups of individuals, but also emerges at multiple scales.

Here we address the issue of hierarchy in citation networks by considering 266 temporal networks, each defined by a \emph{category} or a \emph{keyword} given by Thomson Reuters' \emph{Web of Science} (WoS) database during the period 1975--2011 \cite{wos13}.
From the possible networks we have chosen all samples that are large enough and have an appropriate complexity for our analysis.
Details about the generation of temporal networks are described in the Methods.
We measure the level of flow hierarchy by a corresponding metric, the \emph{global reaching centrality} ($\GRC$) \cite{mones12}:
\begin{equation}
  \GRC=\frac{\sum_i\lrc^M-\lrc(i)}{N-1}=\frac{N}{N-1}(\lrc^M-\langle\lrc\rangle),
  \label{eq:grc_def}
\end{equation}
where $\lrc(i)$ is the fraction of nodes that node $i$ can reach via out-links, $N$ is the number of nodes in the graph.
Here $\lrc^M$ denotes the maximal value and $\langle\lrc\rangle$ is the average.

After determining the behavior of $\GRC$ in the 266 networks, we consider a simple argument about the possible mechanism of the development of hierarchy in citation networks.
The model is based on the assumption that the evolution of hierarchy in these networks is related to the level of generality of the category or keyword.
Finally we support our assumption by showing that the trends of the evolution of the hierarchy changes according to the fraction of external references, i.e., the average number of references to (or from) other fields divided by the number of inner references.

\section{Results}
\label{sec:results}

\subsection{Universal trends}
\begin{figure}[[h!]
  \centering
  \includegraphics[width=0.8\textwidth]{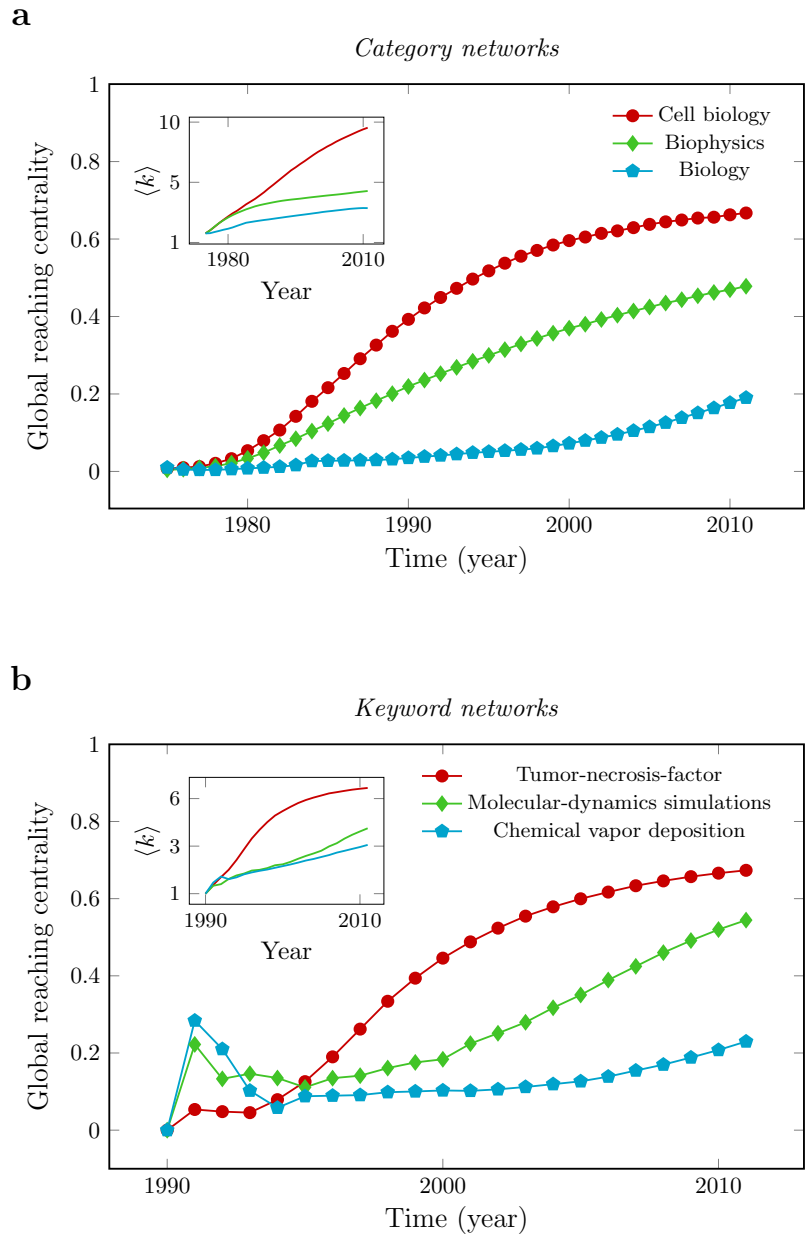}
  \caption{Typical trends of $\GRC$ in the citation networks: (a) category and (b) keyword networks.
Insets show the corresponding average degrees in the considered period.}
  \label{fig:typical_trends}
\end{figure}

We measured the time-dependence of $\GRC$ in 210 category and 56 keyword networks (which we will refer to as \emph{fields} in the rest of the paper, as each category and keyword defines a related field).
Figure~\ref{fig:typical_trends} shows typical trends, together with the evolution of the corresponding average degree.
Note that the corresponding average degree is defined by the number of references, where the cited papers are of the same field.
The first and most striking observation is that the level of hierarchy is increasing in all cases, even though the average degrees corresponding to these fields are also growing (see the inset in Fig.~\ref{fig:typical_trends}).
Studies have found that the extent of hierarchy is quickly vanishing for increasing average degrees in the configuration model \cite{luo11,corominasmutra13,suchecki13}.
This indicates that the evolved structure of the citations is nontrivial, and even networks with low $\GRC$ values are considerably more hierarchical than most of other real-world networks \cite{mones12}.
However, this continuous growth is the fingerprint of these networks, and it is strongly related to the fact that new edges do not emerge between nodes that are already present, but are only added between new nodes and old ones (aside from the negligible fraction of directed loops that are the results of errors in the database in most of the cases).
The primary difference between the curves is the intensity of growth: the networks for category \emph{cell biology} and keyword \emph{tumor-necrosis-factor} show a rapid increase in the global reaching centrality, while the networks related to the other categories and keywords exhibit slower growth.
The fields shown in Fig.~\ref{fig:typical_trends} are representing the typical behaviors: the bottom and top curves represent the slowly and fast growing fields, and the middle trend illustrates those between the two extremals.

\begin{figure}[h!]
  \centering
  \includegraphics[width=0.8\textwidth]{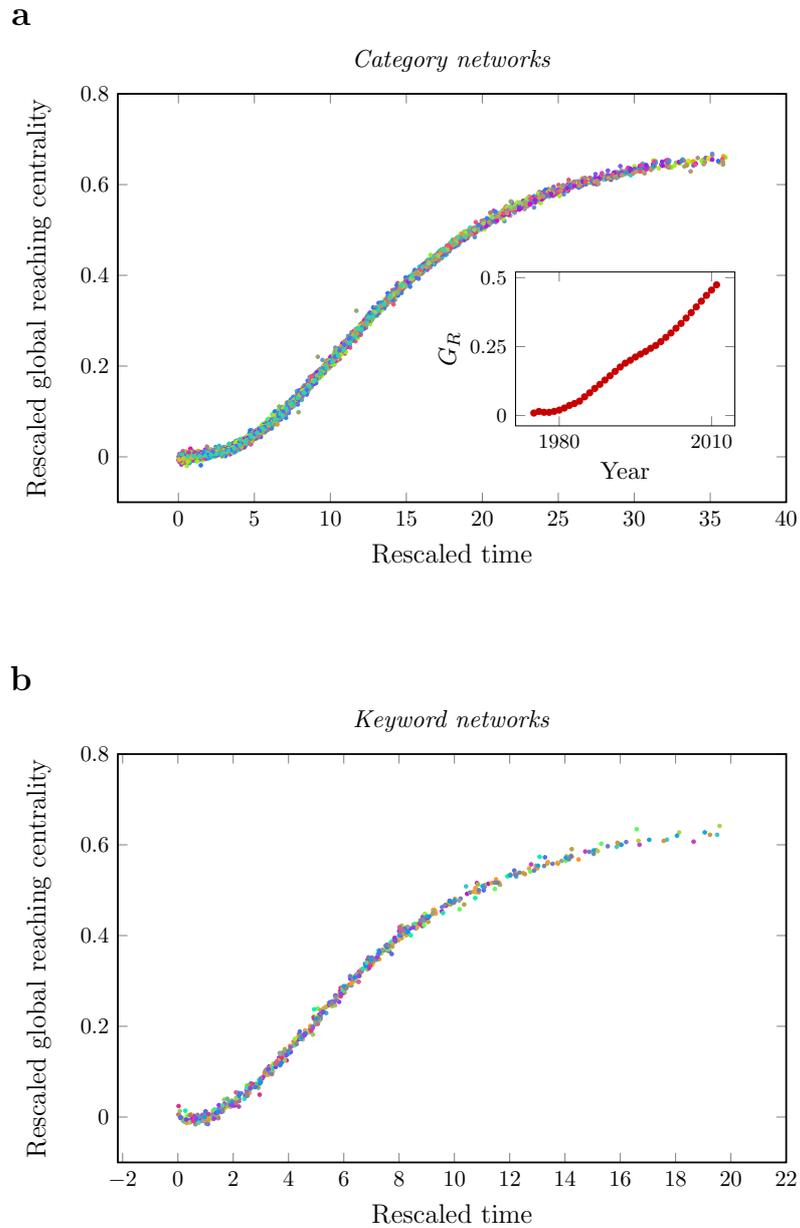}
  \caption{Collapsed trends in the two citation networks: (a) category and (b) keyword networks.
Axes denote the rescaled time (horizontal) and $\GRC$ (vertical) after the linear transformation (translation and scale) of the curves.
The inset in (a) shows a typical curve that was rejected in the rescaling process due to the intermittent increase characteristics (it corresponds to the category \emph{engineering-electrical and electronic}).}
  \label{fig:collapsed_data}
\end{figure}

This similarity in the hierarchical curves suggests that the main mechanisms behind all categories have common features.
This is also supported by Fig.~\ref{fig:collapsed_data}, where we rescaled and plotted all curves to fit the universal trends defined by the category \emph{cell biology} and keyword \emph{tumor-necrosis-factor}.
Details of the rescaling process are presented in the Methods section.
In Fig.~\ref{fig:collapsed_data}, we considered only networks that have a final $\GRC$ larger than 0.1, to ensure that the vertical scaling gives reasonable values (i.e., to obtain less noisy scalings).
It is clearly visible that the hierarchical development of the different fields can be described by a single universal tendency (both for category and keyword networks), and the fields differ mostly in the time-scale.
Citation networks can reach different level of final hierarchy that is specific of the fields, but on the long run, all fields exhibit the same behavior.
Although most of the curves can be fitted on the universal trend by linear transformations, there are 11 category and 7 keyword networks that show a rather different behavior.
However, these networks also show increasing hierarchy with a piecewise sigmoid-like shape (see inset in Fig.~\ref{fig:collapsed_data}, where the category \emph{engineering, electrical and electronic} is plotted).
Also, some of the curves that reach their final state in a few years are continued after the last year shown in the collapsed data, but they remain at a constant value, or grow with a negligible rate, therefore these sections are not shown in the figures.

\subsection{A simple model}
\begin{figure}[h!]
  \centering
  \includegraphics[width=\textwidth]{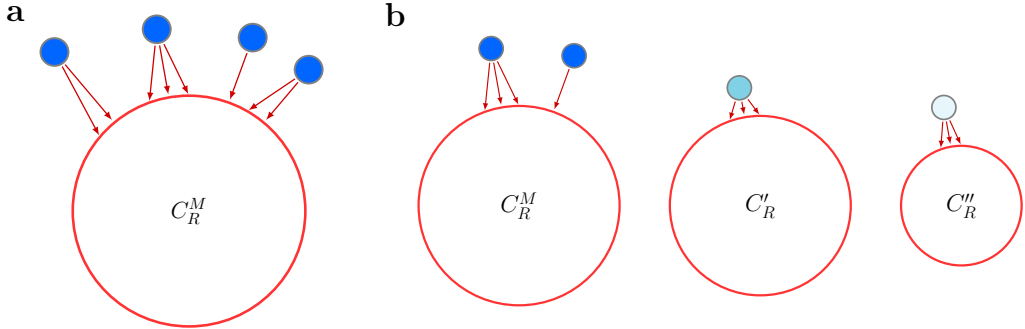}
  \caption{Illustration of graphs with different structures below the nodes with the larges reaching centrality.
In the case of a specialized field (a), the network is growing below a small set of top nodes (blue ones) and they have very similar reaching centralities.
If the field is more general (b), the network is broken down into small reachable sets of the top nodes and they represent the diverse research streams under the common field.
This network type is characterized by a much smaller fraction of nodes with $\lrc$ values close to the maximum one, and also the actual values of the largest ones vanishes more quickly (as they correspond to diverse reachable sets).}
  \label{fig:lrcmax_structure}
\end{figure}

In order to understand the differences seen in the trends, here we present a simple dynamical model for the time evolution of the global reaching centrality.
The model relies on two assumptions about the structure of citation networks.
First, $\GRC$ is related to the difference between the largest and average reaching centralities (see Eq.~(\ref{eq:grc_def})).
As more detailed measurements on the hierarchy reveal, that the observed trends are dominated by the $\lrc^M$ since $\langle\lrc\rangle$ is negligible.
This is rather clear: when new papers connect to the actual network, they have only in-edges (because edges point towards the citing papers).
As the number of papers increases accordingly to an exponential growth \cite{price63,suh09}, the increasing number of nodes without out-links reduces the average reaching centrality to near zero.
The second assumption is less obvious and it can be described as follows.
Each network is defined by a specific field, given by the corresponding category or keyword.
Every field is characterized by several main research streams, which are ignited by a handful of papers for each stream and the rest of the network is growing \emph{below} these standard works.
To illustrate this idea, in Fig.~\ref{fig:lrcmax_structure} we depicted two scenarios corresponding to the different levels of generality of the fields: \emph{specialized} that is more focused on a narrow topic; and \emph{general}, which covers a broader range of topics.
In the case of a more specialized field (Fig.~\ref{fig:lrcmax_structure}a), the whole network is much more interconnected and the papers at the \emph{top} of the hierarchy -- those with the largest reaching centralities -- have more or less the same number of nodes they can reach.
In other words, the nodes with the largest $\lrc$ have similar values, thus more node has a local reaching centrality nearly $\lrc^M$.
On the other hand, in a general field including diverse research streams (Fig.~\ref{fig:lrcmax_structure}b), top papers have varying $\lrc$ values.
Thus, the fraction of nodes having nearly $\lrc^M$ is much less than in the case of a specialized field.

\begin{figure}[h!]
  \centering
  \includegraphics[width=0.8\textwidth]{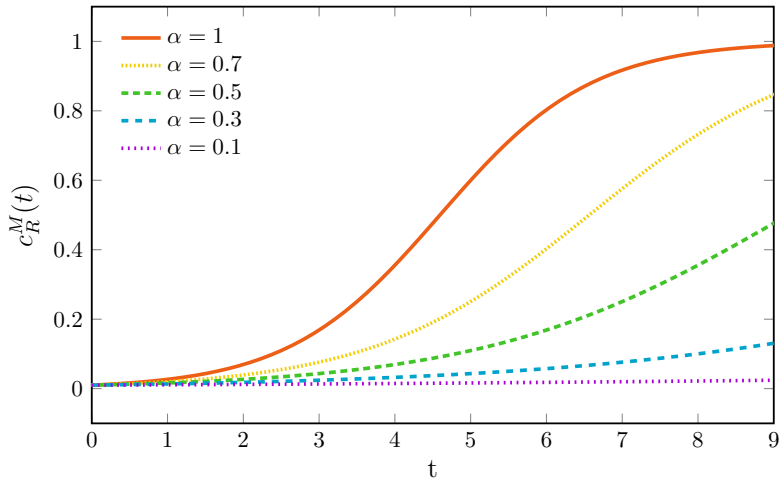}
  \caption{Solutions of Eq.~(\ref{eq:lrcmax_dynamics}) at different values of the parameter $\alpha$.
The initial value is set to $\lrc^0=0.01$ for all curves.}
  \label{fig:lrcmax_solutions}
\end{figure}

Based on the first observation, we will approximate the $\GRC$ by $\lrc^M$ and construct an equation describing its rate of change.
When a new paper appears, it is immediately attached to the network by its references.
The probability that the node is attached to the reachable set of the top node is proportional to the corresponding reaching centrality $\lrc^M$.
It can be shown that the increment of $\lrc^M$ at the attachment of a single node is $(1-\lrc^M)/(N+1)$.
Also, the fraction of nodes with reaching centrality near $\lrc^M$ is related to the generality of the field: if the category or keyword describes a field with few research streams (it is a rather specialized one), the probability of finding a node with a large local reaching centrality is high.
On the contrary, if the field is more general, this probability is reduced.
Assuming that $\alpha N$ number of the nodes have reaching centrality close to $\lrc^M$, where $\alpha$ tunes the generality of the network, these points can be formulated in the following dynamical equation for $\lrc^M$ in the large $N$ limit:
\begin{equation}
  \frac{\diff\lrc^M}{\diff t}=\alpha\,\lrc^M\,(1-\lrc^M),
  \label{eq:lrcmax_dynamics}
\end{equation}
The above logistic equation has the well-known solution
\begin{equation}
  \lrc^M(t)=\frac{1}{1+\Big(\frac{1}{\lrc^0}-1\Big)e^{-\alpha t}},
  \label{eq:lrcmax_solution}
\end{equation}
which describes a sigmoid with $\lrc^0$ as its initial value.
To illustrate the solution, we plotted Eq.~(\ref{eq:lrcmax_solution}) at $\lrc^0=0.01$ and different values of the parameter $\alpha$ in Fig.~\ref{fig:lrcmax_solutions}.
As we can see, varying the parameter $\alpha$, the solution gives the typical trends seen in Fig.~\ref{fig:typical_trends}, considering a finite interval of time.
Figure~\ref{fig:lrcmax_solutions} suggests that the differences among the categories and keywords might be originated in the different level of generality, i.e., in the number of research streams they cover.

\subsection{Separation of trends}
\begin{figure}[h!]
  \centering
  \includegraphics[width=0.8\textwidth]{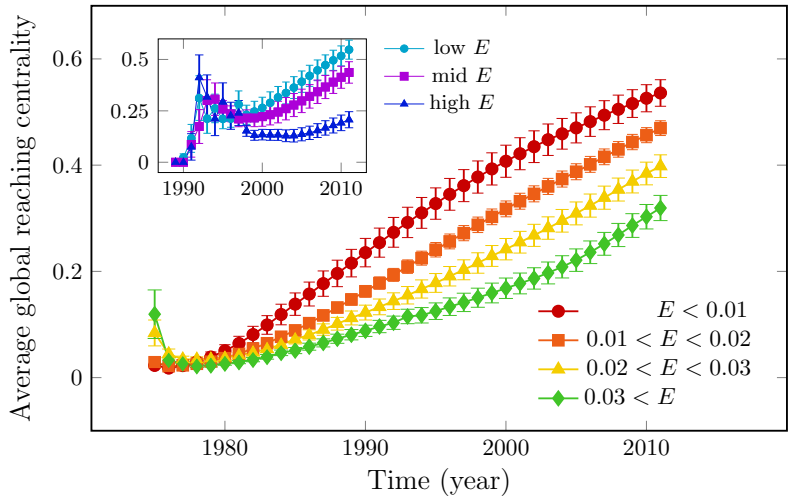}
  \caption{Hierarchical trends in the category networks with different external reference ratio ($E$).
The curves correspond to the averages taken over the networks with the indicated value of $E$.
Trends of the keyword networks are shown in the inset: in this case, they are separated into 3 groups of equal number of networks according to the values of $E$.
Error bars denote the standard deviation of the mean.}
  \label{fig:grouped_trends}
\end{figure}

Motivated by the model, now we take a closer look at the curves seen in Figs.~\ref{fig:typical_trends}-\ref{fig:collapsed_data}.
In order to check the effect of generality on the time evolution trend of hierarchy, we considered the cross-category references among the papers in the category networks.
More precisely, for each category $\mathcal{C}$, we calculated the number of references ($R_{\mathcal{C},\mathcal{C}}$) between pairs of papers, both labelled by the same category $\mathcal{C}$.
Similarly, we calculated the average number of references ($\langle R_{\mathcal{C},\mathcal{C'}}\rangle_{\mathcal{C'}\ne\mathcal{C}}$) between papers of a given category $\mathcal{C}$ and papers of other categories.
These two quantities describe the number of \emph{internal} ($R_{\mathcal{C},\mathcal{C}}$) and \emph{external} ($\langle R_{\mathcal{C},\mathcal{C'}}\rangle_{\mathcal{C'}\ne\mathcal{C}}$) references of category $\mathcal{C}$.
Considering the generality of the category, we calculated the \emph{external reference ratio}:
\begin{equation}
  E(\mathcal{C})=\frac{\langle R_{\mathcal{C},\mathcal{C'}}\rangle_{\mathcal{C'}\ne\mathcal{C}}}{R_{\mathcal{C},\mathcal{C}}}.
  \label{eq:err_def}
\end{equation}
As an approximation of the problem of generality, we use this ratio to quantify how specialized a category is: if $E(\mathcal{C})$ is very small, papers in the category $\mathcal{C}$ prefer to cite papers inside their own category, meaning that it is a rather specialized field.
However, if the category covers many distinct subfields, it implies a higher fraction of external citations (and vice verse, papers in other fields cite the category papers more frequently), resulting in a larger value of $E$.
In the 210 categories under consideration, the value of $E(\mathcal{C})$ lies in the range $[0:0.1]$.
We divided the networks into four groups by their external reference ratio, so that each group contains at least 27 categories, and calculated the average $\GRC$ trends inside the groups.
The results are plotted in Fig.~\ref{fig:grouped_trends}, showing well-separated hierarchical curves.
The curves confirm the assumption of the model: the more specialized a category is, the sooner it develops a highly hierarchical structure.
The inset of Fig.~\ref{fig:grouped_trends} shows the same curves for the keyword networks.
Due to the small number of inner edges in these networks, the data are very noisy at the beginning, but the trends in later years are better separated.

\begin{figure}[h!]
  \centering
  \includegraphics[width=0.8\textwidth]{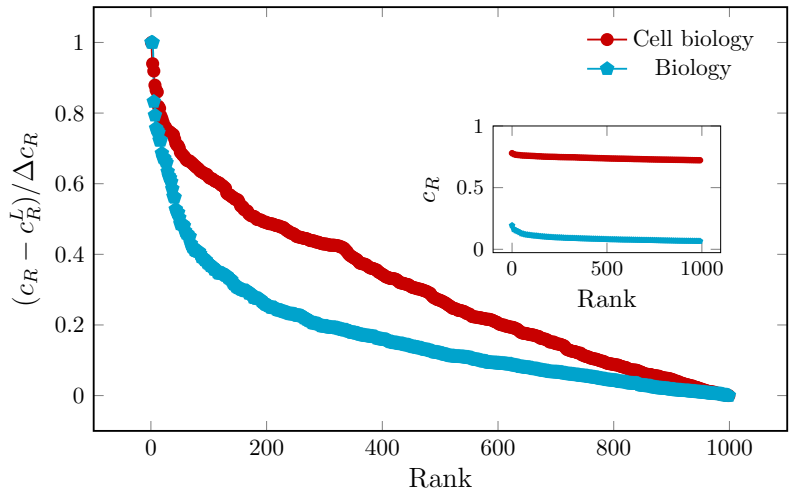}
  \caption{Largest 1000 ranked $\lrc$ values in the year 2011 for two category networks, sorted in descending order.
Horizontal axis denotes the rank of $\lrc$, starting with the maximum one at rank zero.
On the vertical axis, the relative value of $\lrc$ is plotted, normalized in a way that the offset ($\lrc^L$ which is the centrality at rank 1000) is subtracted and the data is divided by the vertical extent ($\Delta\lrc=\lrc^M-\lrc^L$) of the curves.
The original curves are shown in the inset.}
  \label{fig:lrctop}
\end{figure}

As an additional corroboration, we can calculate the values of $\lrc$, to test whether more general categories are characterized by fewer nodes with near-maximal reaching centrality.
For this, in Fig.~\ref{fig:lrctop} we plotted the first 1000 largest $\lrc$ values against their ranks in the categories \emph{cell biology} and \emph{biology}, measured in the final state of the networks.
To emphasize the difference between the two categories, we transformed the curves to a form that is easier to grasp, and we also provide the original data.
The abscissa shows the rank of the $\lrc$ value, starting with the largest one.
On the ordinate, the normalized values are shown: from each value, we subtracted the last (i.e., the 1000th) data point ($\lrc^L$), and divided by the difference between the two extremal values ($\Delta\lrc=\lrc^M-\lrc^L$).
Clearly, the values in the category \emph{biology} (having a larger value of $E$) vanish faster, meaning that the probability of having near-maximal reaching centrality is very low, compared to the category \emph{cell biology}.
Thus, according to the external reference ratio, the assumption of the model about the fraction of nodes with reachable set similar to the largest one is reasonable and shows good agreement with the data.

\section{Methods}
\label{sec:methods}

\subsection{Temporal networks}
Two types of citation networks are considered in this paper:
\begin{itemize}
  \item[]\emph{Category networks}: They are defined by the Subject Categories given by the WoS database.
Nodes of these networks are papers corresponding to one Subject Category, and an edge is linked from paper $A$ to paper $B$ if paper $B$ cited paper $A$.
  \item[]\emph{Keyword networks}: These are based on specific keywords: all nodes of a keyword network are papers that are labeled, inter alia, by a selected keyword.
As with the category networks, edges represent citation and are directed so that they point from the cited paper towards the citing paper.
\end{itemize}
The reason behind the above definition of edge directions is that we are interested in the network of influences among papers, and we assume that papers can influence only later papers.
For each category, we generated temporal networks in the period $y=1975,\dots,2011$, by aggregating the networks from year 1975 to year $y$.
There are 251 Subject Categories in total, each paper in the database is labeled at least with one category.
As an example, three categories are \emph{acoustics}, \emph{geography} and \emph{robotics}.
In order to detect whether there is a possible bias caused by the fact that papers in the WoS database are indexed only from 1975, in the case of keyword networks, we only consider keywords that were born between years 1990 and 1995.
These keywords are defined as having a frequency of appearance 10 times larger in 1995 than in 1990 and still being alive in 2011 by appearing at least in 1000 papers.

\subsection{Data filtering}
Several filtering processes were applied to the data, due to the presence of categories and keywords with very small networks and very low connectivity.
To decrease the level of noise, we first rejected networks with a final size less than 1000.
In other words, fields with at least 1000 papers in total are considered.
Furthermore, we also rejected networks with a final average degree lower than 1, because these networks consist of disjoint small components and they cannot be characterized by a coherent structure.
This is crucial, since we are aiming at the study of the evolution of a field as a whole, which is not true for the rejected networks, as any measured property is merely the mixture of the independent subgraphs (smaller subfields).
Similarly, we also set a threshold on the size of the giant components: they must have a \emph{giant weakly connected component} (GWCC) larger than 0.5, for the reason mentioned before.
The GWCC is defined by the largest connected component of the graph.
After the above filters, 210 category and 56 keyword networks remained.

\subsection{Hierarchy}
\begin{figure}[h!]
  \centering
  \includegraphics[width=0.8\textwidth]{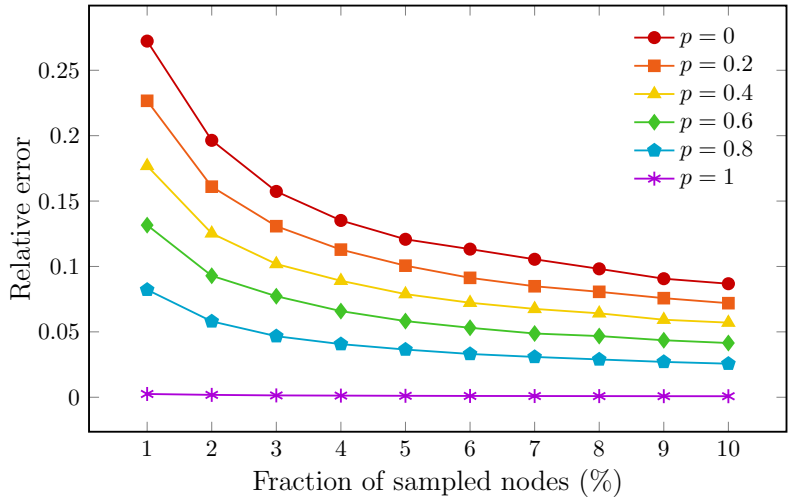}
  \caption{ Relative errors of the estimated $\GRC$ in a synthetic model at different hierarchy parameter values.
Description of the synthetic model can be found in Ref.\cite{mones12}.
}
  \label{fig:grc_estimate}
\end{figure}

The measurement of the flow hierarchy in the citation networks is based on the global reaching centrality ($\GRC$) given by Eq.~(\ref{eq:grc_def}).
For this measure, one is supposed to determine the largest and the average local reaching centralities, which requires $\mathcal{O}(N^2)$ operations using a simple depth-first search algorithm.
In our case, the size of the networks increases quickly, and it is required to calculate the $\GRC$ for graphs with $N\approx10^6$ nodes and $N\approx10^7$ edges.
Therefore, we applied an approximating method for the calculation of the hierarchy measure.
The method is divided into the following steps:
\begin{enumerate}
  \item Choose a node randomly.
  \item Calculate the local reaching centrality $c_R(i)$ for the chosen node.
  \item Update $c_R^M$ and if the node was chosen randomly, update $\langle c_R\rangle$ as well.
  \item With probability $\pjump$ go to Step 1 (if $k_{\mathrm{in},i}=0$, set $\pjump=1$ for this step).
  \item With probability $1-\pjump$, choose an in-neighbor randomly and go to Step 2.
\end{enumerate}
On the one hand, nodes are chosen randomly for $\pjump$ fraction of the steps.
Since these node samples are uncorrelated, the approximation of the average is unbiased.
On the other hand, in $1-\pjump$ fraction of the steps, the method walks through in-edges, and therefore discovers nodes at higher levels of the hierarchy.
The average relative errors of the estimations defined by
\begin{equation}
  \langle\delta\GRC\rangle=\bigg\langle\frac{\GRC^\mathrm{exact}-\GRC^\mathrm{estimated}}{\GRC^\mathrm{exact}}\bigg\rangle
  \label{eq:rel_err_def}
\end{equation}
are shown in Fig.~\ref{fig:grc_estimate}, where we calculated $\langle\delta\GRC\rangle$ for a hierarchical synthetic model that is introduced in \cite{mones12}.
The synthetic model is controlled by a parameter $p$, which tunes the extent of hierarchy.
The networks have $N=10000$ nodes and $\langle k\rangle=3$ in each case.
All dots are the average of 100 different samplings and also 100 different network realizations at $\pjump=0.1$.
Although the method has rather large errors at low hierarchy ($>10\%$), as the network becomes more hierarchical, the error tends to decrease quickly below an admissible level.
However, it should be noted that networks with negligible extent of hierarchy have a value of $\GRC<<1$ and therefore any conclusions drawn from its value remains valid even at an error of $10\%$.
In the calculations for real networks, we applied $\pjump=0.1$ and a number of jumps being equal to 10000.

\subsection{Data collapse}
The collapsed data shown in Fig.~\ref{fig:collapsed_data} are obtained by the rescaling of the trends measured in the various fields.
As Fig.~\ref{fig:typical_trends}b shows, when the networks consist of very few papers (less than 100), the global reaching centrality can take large values as well, which later vanish as the networks grow to large enough.
Therefore, we considered the $\GRC(t)$ trends only from the year, at which the networks have at least 100 nodes.
Moreover, some of the networks gain a finite value of $\GRC$ in the beginning and this value is not changing over the next few years, since there is no significant change in the structure.
Thus, we also removed the offset in the networks by subtracting $\GRC(t=0)$ from all other years.
We considered linear transformations of the trends in order to rescale them onto the base trends which are one of the most articulated ones (the networks related to \emph{cell biology} and \emph{tumor-necrosis-factor}).
The transformation of the trend $\mathcal{T}$ onto the base trend $\mathcal{T}_\mathrm{base}$ was carried out by minimizing the function $\mathcal{H}=\langle R^2\rangle/L$, where $\langle R^2\rangle$ denotes the average squared difference of data $\mathcal{T}_\mathrm{base}$ and $\mathcal{T}$ in the overlapping region of the two data, and $L$ is the length of the overlap.
We optimized $\mathcal{H}$ by simulated annealing with respect to horizontal and vertical translations and scalings, and also to the removal of some of the first data points (5 of them at most).

\section{Discussion}
\label{sec:discussion}

In this paper we have studied the evolution of the hierarchical structure of citation networks restricted to categories and keywords given by the \emph{Web of Science} database.
We measured the time-dependence of flow hierarchy in the networks by a corresponding metric, and surprisingly found that a majority of the categories follow a unique growth tendency.
It is known that hierarchical structures are less likely to appear and be present in networks with large average degree \cite{luo11,corominasmutra13,suchecki13}.
Although the growing citation networks are characterized by an increasing average degree, they also form a more hierarchical structure in time.
The nature of hierarchy in the networks is further supported by the fact that most of the curves can be collapsed into one universal curve.
To investigate the observed tendency, we proposed a simple model that is able to capture the main qualitative properties.
By a few assumptions, we showed that the different pace of the hierarchy tendencies is related to the specialization level of the corresponding field (category or keyword).
It is confirmed by a simple model with the aim of characterizing the dynamics of the hierarchy in the networks, and also through the separation of the data by a quantity that is based on additional information of the papers and captures the generality of the related categories.

Finally, a last comment should be added to the model and especially Eq.~(\ref{eq:lrcmax_dynamics}).
Obviously, the true dynamics in real networks should not be as simple as we described it, and the purpose of the model is to indicate that the generality of the defining field of the network has an impact on the hierarchy.
In this setup, when a new paper is born, it has multiple references, and thus the right hand side of Eq.~(\ref{eq:lrcmax_dynamics}) should be multiplied by the average degree $\langle k\rangle$.
Implicitly, in the derivation of the dynamical equation, we assumed that the only parameter $\alpha$ includes all further multiplicative factors, and therefore it counts for $\langle k\rangle$ as well.
Furthermore, the solution given by Eq.~(\ref{eq:lrcmax_solution}) has an asymptotic value of $\lrc^M=1$, independently of the initial value or the parameter $\alpha$, in contrast to the real networks.
This is related to the fact that real networks do not have the exact structural form described in the model: they are not simple tree-like subgraphs melted together, but much more complex objects, having nontrivial -- and not even layered -- inner structure.
Thus, the number of nodes not living in the reachable set of the top node is not exactly $(1-\lrc^M)N$, but somewhat lower, resulting in a smaller final level of hierarchy.

The growing level of hierarchy found in citation networks suggests that papers are organized in a highly ordered structure.
As a hierarchy, it is characterized by a small fraction of papers that dominate the network through the citations they receive.
Papers at the top of the hierarchy are also the root of smaller subgraphs (research streams), and as the network grows, these subgraphs merge together.
Although the hierarchical structure can be understood by simple models, there are open questions remaining.
What are the special characteristics of papers at the top of the hierarchy?
How do top papers evolve in time, and how is the success of a publication related to its position in the hierarchy?
How does the structure of the neighborhood of a paper correlate with its success?
As a further investigation, these questions can be addressed by a more detailed model and exploration of the local properties of citations.

\section{Acknowledgments}
\label{sec:acknowledgements}
This work was partially supported by the European Union and the European Social Fund through project FuturICT.hu (Grant No: T\'{A}MOP-4.2.2.C-11/1/KONV-2012-0013), by the Hungarian National Science Fund (OTKA K105447), and by the EU ERC FP7 COLLMOT Grant No: 227878.


\end{document}